\documentclass[10pt,conference]{IEEEtran}
\IEEEoverridecommandlockouts
\usepackage[T1]{fontenc} % issue with Verbatim fonts
\usepackage{microtype}
\usepackage{amsmath,amssymb,amsfonts}
\usepackage{algorithm}
\usepackage{algorithmic}
\usepackage{graphicx}
\usepackage{textcomp}
\usepackage{xcolor}
\usepackage{cite}
\usepackage{orcidlink}
\usepackage{booktabs}
\usepackage{multirow}
\newlength{\todowidth}
\usepackage{xspace}
\usepackage{fancyvrb}
\usepackage{siunitx}
\usepackage{paralist}
\usepackage{comment}
\usepackage{hyperref}
\includecomment{draftversion}
\excludecomment{submittedversion}

\begin{document}
%%%%%%%%%%%%%%%%%%%%%%%%%%%%%%%%%%%
% \title{AI-based Dynamic Management of a Parallel Skeleton over the OpenFaaS Serverless Platform  (provisional title)

% \title{Reinforcement Learning–Based Autoscaling of a Farm Skeleton on the OpenFaaS Serverless Platform
\begin{draftversion}
\title{Reinforcement Learning–Based Dynamic Management of Structured Parallel Farm Skeletons on Serverless Platforms
\thanks{Corresponding author: Lanpei Li --- This research was partially supported by the FIS2 Grant from Italian Ministry of University and Research (Grant ID: FIS2023-03382).}}
\author{
    \IEEEauthorblockN{
        Lanpei Li $^{\ast,\dagger}$ \orcidlink{0009-0005-4370-1020},
        Massimo Coppola $^{\ast}$ \orcidlink{0000-0002-7937-4157},
        Malio Li $^{\dagger}$ \orcidlink{0009-0002-9572-7546},
        Valerio Besozzi $^{\dagger}$ \orcidlink{0009-0002-8493-2122},
        Jack Bell $^{\dagger}$ \orcidlink{0009-0002-7857-4752},
        Vincenzo Lomonaco $^{\ddagger}$ \orcidlink{0000-0001-8308-6599},
    }
    \IEEEauthorblockA{
        ${\ast}$\textit{Institute of Information Science and Technologies (ISTI), National Research Council of Italy (CNR)}, Pisa, Italy \\
        lanpei.li@isti.cnr.it, massimo.coppola@isti.cnr.it
    }
    \IEEEauthorblockA{
        ${\dagger}$\textit{Department of Computer Science, University of Pisa}, Pisa, Italy \\
        malio.li@phd.unipi.it, valerio.besozzi@phd.unipi.it,
        jack.bell@di.unipi.it
    }
    \IEEEauthorblockA{
        ${\ddagger}$\textit{Department of AI, Data and Decision Sciences, LUISS University}, Rome, Italy \\
        vlomonaco@luiss.it
    }
}

\end{draftversion}

\begin{submittedversion}
\title{Reinforcement Learning–Based Dynamic Management of Structured Parallel Farm Skeletons on Serverless Platforms
\thanks{Corresponding author: \textit{(omitted)} --- This work has been partially supported by the  project \textit{(omitted)} under grant ID \textit{(omitted)}.}}
% \author{---, ---, ---, ---, ---, ---}
\author{
\IEEEauthorblockN{Anonymized Authors}
\IEEEauthorblockA{\textit{Anonymized Affiliations} \\
\textit{Anonymized City, Anonymized Country} \\
email@anonymized.com}
}
\end{submittedversion}

%\date{July 2025}
%%%%%%%%%%%%%%%%%%%%%%%%%%%%%%%%%%%
\begin{submittedversion}
% \title{AI-based Dynamic Management of a Parallel Skeleton over the OpenFaaS Serverless Platform  (provisional title)

\title{Reinforcement Learning–Based Dynamic Management of Structured Parallel Farm Skeletons on Serverless Platforms}

% \title{Reinforcement Learning–Based Autoscaling of a Farm Skeleton on the OpenFaaS Serverless Platform}

% \author{anon1, anon2, anon3, anon4, anon5, anon6}
\author{
\IEEEauthorblockN{Anonymized Authors}
\IEEEauthorblockA{\textit{Anonymized Affiliations} \\
\textit{Anonymized City, Anonymized Country} \\
email@anonymized.com}
}
\end{submittedversion}

%%%%%%%%%%%%%%%%%%%%%%%%%%%%%%%%%%%

\maketitle
\begin{abstract}
%\emph{see old abstract in {\tt OLD\_main.tex} --- revised to refocus only on the farm}\\
%
We present a framework for dynamic management of structured parallel processing skeletons on serverless platforms. Our goal is to bring HPC-like performance and resilience to serverless and continuum environments while preserving the programmability benefits of skeletons. As a first step, we focus on the well known \textit{Farm} pattern and its implementation on the open-source OpenFaaS platform, treating autoscaling of the worker pool as a QoS-aware resource management problem. %facilitate their composition into applications and 

The framework couples a reusable farm template with a Gymnasium-based monitoring and control layer that exposes queue, timing, and QoS metrics to both reactive and learning-based controllers. We investigate the effectiveness of AI-driven dynamic scaling for managing the farm's degree of parallelism via the scalability of serverless functions on OpenFaaS.
%
%While 
%the framework we envise shall support all well-established parallel skeletons, 
% We focus on the well known \textit{Farm} pattern, and its implementation on top of OpenFaaS
%, and the definition of its dynamic resource management problem in such settings. 
%
% Our prototype employs AI-driven dynamic scaling to overcome OpenFaaS limitations. We discuss the autoscaling model and its training, and evaluate two RL policies against a baseline of reactive management based on the Farm performance model.
%
In particular, we discuss the autoscaling model and its training, and evaluate two reinforcement learning (RL) policies against a baseline of reactive management derived from a simple farm performance model.
%models when training and enacting informed decision-making.
%
Our results show that AI-based management can better accommodate platform-specific limitations than purely model-based performance steering, improving QoS while maintaining efficient resource usage and stable scaling behaviour.

%. The approach involves examining well-established parallel patterns such as Pipeline, Farm, and Loop, creating the fundamental abstractions for structured parallelism and supporting their composition. 
%investigate improving its serverless-based skeleton support by adding AI-based management, targeting performance, efficiency and QoS enforcement.
%
%
%
%This work focuses on the Farm skeleton, its implementation on top of OpenFaaS, and the definition of its dynamic resource management problem. We propose a solution based on AI dynamic scaling and evaluate it against a baseline policy employing reactive management and the skeleton's performance-model.

%\todo{Additionally, the framework introduces a Quality-of-Service (QoS)-aware parallel control API explicitly designed for reinforcement learning (RL)-based dynamic and autonomic control.}\sidenote{MC: how much to stress AI in this first work?}
%
%Standardized templates and dynamic mappings of user-defined computational tasks to function-as-a-service (FaaS) instances enhance resource efficiency and adaptability. 

\end{abstract}

\section{Introduction}
Serverless Function-as-a-Service (FaaS) platforms promise elastic scaling and operational simplicity, but their black-box autoscalers and cold-start behaviour make it hard to provide performance guarantees for latency-sensitive parallel workloads~\cite{manner2022faasScaling,agarwal2024drl}. In contrast, structured parallel programming frameworks expose high-level skeletons---such as the farm pattern---that decouple application logic from low-level parallel execution and offer a small set of control knobs, most notably the degree of parallelism~\cite{aldinucci2017guide}. Bridging these two worlds requires runtime mechanisms that expose skeleton-level control while coping with platform-specific limitations and overheads.  
%danelutto2014structured,

In this work we study the dynamic management of a task farm skeleton deployed on the open-source OpenFaaS\footnote{See \url{https://www.openfaas.com/}} platform. The farm processes a continuous stream of image-processing tasks under explicit deadline-based quality-of-service (QoS) objectives. We design an OpenFaaS-based farm template, calibrate a workload and QoS model from sequential executions of the image-processing pipeline, and investigate autoscaling policies that adjust the farm's worker pool size over time in response to non-stationary load, building on and extending prior work on serverless task farming~\cite{kehrer2019serveless,SelfTuning-Kehrer2021}. 

We contrast analytical reactive baselines derived from a behavioural model of the farm with RL agents that learn scaling policies directly from interaction with the system via a Gymnasium-style %environment~\cite{sutton2018reinforcement,tesauro2006hybrid,mao2016resource,agarwal2024drl,towers2024gymnasium,gari2021reinforcement,qiu2023aware}.   % why this bunch of references here?
environment~\cite{towers2024gymnasium}. 
Our goal is to understand when AI-based control can compensate for platform-induced delays and non-ideal behaviours that are hard to capture analytically, and how it compares against simple reactive strategies in terms of QoS, efficiency, and stability. Our current evaluation focuses on a single tenant scenario and a single skeleton instance; more complex settings
%multi-tenant interference and interactions among multiple skeleton instances 
are left to future work.
The skeleton, simulation and policy code is open source\footnote{Source code repository: \url{https://github.com/lilanpei/RL-OpenFaaS-Farm}}.

This paper makes the following contributions:
\begin{inparaenum}[(1)]
\item 
Design and implementation of a Farm parallel skeleton (as a specific instance of a Parallel Processing Skeleton) running on the OpenFaaS serverless platform, with Emitter, Worker, and Collector functions connected via Redis-backed queues.
\item
Formalisation of a QoS-aware autoscaling problem for the farm skeleton, including a workload and deadline model and a set of metrics capturing performance, efficiency, and QoS.
\item
An AI-based dynamic management approach that complements analytical reactive baselines with RL policies aimed at optimising a combination of QoS, resource usage, and scaling stability.
\item
Design of the farm autoscaling environment with a set of Gymnasium-style~\cite{towers2024gymnasium} run-time monitoring hooks that expose farm-level state, actions, and rewards for both analytical and learning-based controllers.
\item
An experimental comparison on a workload calibrated for an image-processing pipeline. We evaluate reactive and RL-based policies on top of OpenFaaS in terms of QoS compliance, queue dynamics, and worker utilisation.
\end{inparaenum}

%\todo{Paper structure: TODO}
The remainder of the paper is organised as follows. Section~\ref{sec:background} 
%reviews background on serverless task farms, AI-based autoscaling, reinforcement-learning methods, and Gymnasium-style environments. 
positions our work in the context of prior research, briefly reviewing serverless task farms, AI-based autoscaling for Cloud and FaaS platforms, and the reinforcement-learning methods that underpin our methodology.
Section~\ref{sec:openfaas:skeletons} summarizes the options and constraints the OpenFaaS platform provides to our design.
%deployment and farm implementation.
Section~\ref{sec:formulation} formalises the QoS-aware autoscaling problem for the farm skeleton.  Section~\ref{sec:farm:implem} presents reactive and RL-based management policies. Section~\ref{sec:experimental-settings} details the experimental setup and evaluation, and Section~\ref{sec:conclusions} concludes with a summary and future work directions.

\section{Background}\label{sec:background}
The concept of algorithmic skeletons was first introduced and popularized by Murray~\cite{cole1989algorithmic}. The term \textit{skeleton} refers to the underlying structure or pattern of a parallel algorithm that can be reused across different domains~\cite{history-spp}. 
Algorithmic skeletons are essentially generic, adaptable, and reusable building blocks that can be used to parallelize an application.
%They help implementing efficient parallel solutions with an abstract, high-level approach, removing the need to manage low-level parallel details~\cite{kessler2007models}.

%Derived from higher-order functions, commonly known in functional programming, algorithmic skeletons abstract the complexity of parallelism, making it easier to implement efficient parallel solutions without needing to manage low-level parallel details~\cite{kessler2007models}.

%In skeletal parallel programming, programmers focus on expressing the high-level structure and behavior of a parallel algorithm using algorithmic skeletons.
Algorithmic skeletons can be broken down into two parts: the semantics of the skeleton and its implementation~\cite{MCCOOL20121,10.5555/525697.826754}, usually called a \textit{template}. 

%The former describes how an identified parallel pattern can be used as a building block for a parallel algorithm, defining task and data dependencies.
This separation of concerns explicitly hides implementation details to the application developer, enabling high-level abstraction and portability across different systems and architectures. The actual implementation as a template is the responsibility of the skeleton framework developer~\cite{fang_parallel_2020, gonzalezvelez2010survey}, who ensures
%
%On different architectures, the implementation of a specific skeleton may vary, but its semantics remain consistent. 
%This allows developers to focus on choosing the appropriate skeleton (or skeleton composition), rather than planning low-level parallel mechanisms. 
%Ensuring 
the correctness of the parallelism mechanisms behind the skeletons' templates as well as of their composition properties
%
%The latter concept is known as 
(\textit{``parallelism exploitation correctness `by construction'{}''}~\cite{danelutto-mencagli}).

\smallskip
%\subsection{Task Farm Skeleton over Serverless}
Recent work has started to investigate skeleton-style task farming on serverless platforms. Kehrer et al.~\cite{kehrer2019serveless} introduce serverless skeletons for elastic parallel processing, mapping structured patterns such as task farms onto FaaS backends. Building on this line, Kehrer et al.~\cite{SelfTuning-Kehrer2021} propose a self-tuning serverless task farming framework with a proactive elasticity controller that adapts parallelism to meet user-defined execution time limits while minimizing cost. Their architectures target parallel task farms and demonstrate that serverless platforms can support elastic parallel patterns when complemented with domain-specific control logic. Complementary benchmarking work by Manner and Wirtz~\cite{manner2022faasScaling} compares resource scaling strategies for open-source FaaS platforms against commercial cloud offerings, highlighting how platform-level scaling choices affect performance and cost.

Our work differs from~\cite{kehrer2019serveless,SelfTuning-Kehrer2021,manner2022faasScaling} in several aspects. First, we focus on a structured farm skeleton with explicit QoS guarantees for latency-sensitive stream processing, rather than batch task farms or generic cloud functions. Second, we operate on top of OpenFaaS and expose a Gymnasium-compatible environment for experimentation with different control policies. Third, instead of relying solely on model-based or heuristic controllers, we explore RL agents that learn autoscaling policies directly from interaction with the environment, while still allowing comparison against reactive, performance-model-based baselines.

\smallskip
%\subsection{AI-Based Autoscaling and Reinforcement Learning for Resource Management}
AI techniques, and RL in particular, have emerged as promising approaches for cloud autoscaling and resource management. RL agents learn policies that map observed system states to scaling actions in order to optimize long-term reward, which can encode QoS, cost, and stability objectives~\cite{sutton2018reinforcement}. Compared to purely rule-based or control-theoretic approaches, RL can adapt to complex, partially unknown dynamics and non-stationary workloads.

RL has been applied to server provisioning and autoscaling in various settings, including data centres and cloud platforms~\cite{tesauro2006hybrid, mao2016resource,agarwal2024drl}. Concrete RL-based autoscalers have been proposed for Infrastructure-as-a-Service (IaaS) clouds and containerized applications~\cite{barrettApplyingReinforcementLearning2013,khaleqIntelligentAutoscalingMicroservices2021,santosGwydionEfficientAutoscaling2025}. These works demonstrate that RL-based controllers can outperform static or threshold-based policies by anticipating load changes and optimizing multi-objective trade-offs. Agarwal et al.~\cite{agarwal2024drl}, for instance, propose a deep reinforcement learning method for autoscaling FaaS functions that jointly considers QoS and cost in a serverless setting. A recent survey~\cite{gari2021reinforcement} and production learning-based autoscaling systems deployed in practice~\cite{qiu2023aware,rzadcaAutopilotWorkloadAutoscaling2020} further underscore the maturity and practical relevance of learning-based autoscaling techniques. %~\cite{qiu2023aware} further underscore the maturity and practical relevance of RL-based autoscaling techniques. Complementary industrial systems such as Google's Autopilot~\cite{rzadcaAutopilotWorkloadAutoscaling2020} and Alibaba's MagicScaler~\cite{panMagicScalerUncertaintyAwarePredictive2023} illustrate how advanced autoscaling strategies are already being deployed at cloud scale, but without focusing on structured parallel skeletons or serverless farms.
However, many existing approaches target virtual machines, generic microservices, or individual functions and do not consider the specific semantics of parallel skeletons deployed on serverless FaaS platforms. In this work, we position our RL-based autoscaler in this line of research, but tailor the state representation, action space, and reward shaping to the structured farm skeleton and its QoS constraints.

\smallskip
%\subsection{\texorpdfstring
%    {Reinforcement Learning Methods: SARSA($\lambda$) and Deep Q-Networks}
%    {Reinforcement Learning Methods: SARSA(lambda) and Deep Q-Networks}
%}
We adopt two standard value-based RL methods for the farm autoscaling problem, SARSA with eligibility traces and a Double Deep Q-Network (Double DQN)~\cite{sutton2018reinforcement,mnih2015human,van2016deep}, 
and apply them on a common Gymnasium-style environment interface; their concrete instantiation is detailed in Section~\ref{sec:farm:implem}.

When training and comparing different AI techniques over a common experiment, it is essential to decouple environment dynamics from learning algorithms to ease code maintenance and support reproducible studies, so we follow the Gymnasium interface standard~\cite{towers2024gymnasium} for the environment API, which exposes observation and action spaces and allows different RL agents and analytical baselines to be plugged in with minimal changes to the underlying farm deployment.

\begin{comment}
\subsection{Gymnasium Environments for Reinforcement Learning}
OpenAI Gymnasium~\cite{towers2024gymnasium} popularized a simple, standardized interface for RL environments. In the Gymnasium API, an environment exposes:
\begin{itemize}
    \item an observation space (e.g., a vector of real-valued features),
    \item an action space (e.g., a discrete set of scaling actions),
    \item a \texttt{step} function that applies an action and returns the next observation, scalar reward, and a termination flag,
    \item and a \texttt{reset} function that initializes a new episode.
\end{itemize}
This interface decouples environment dynamics from learning algorithms and facilitates reproducible experimentation and benchmarking.

Our OpenFaaS farm environment follows this Gymnasium-compatible design (see Section~\ref{subsec:env}). Exposing the farm skeleton through a Gymnasium-style environment allows us to plug in different RL agents (e.g., SARSA($\lambda$), DQN) and reactive baselines while keeping the underlying deployment and monitoring infrastructure unchanged.
\end{comment}

\section{Parallel Skeletons and the OpenFaaS Serverless Platform}\label{sec:openfaas:skeletons}
%\todo{The OpenFaaS serverless platform ---
%High-level issues in using serverless and specifically OpenFaaS}
%
%
FaaS platforms like OpenFaaS provide a natural runtime for deploying parallel skeleton stages as independent, serverless functions. OpenFaaS functions are packaged as lightweight Docker containers (function pods) that can be deployed on any Kubernetes-based infrastructure. The platform handles function invocation, scaling, and isolation, freeing developers from managing server processes. This suits our goal of making skeleton instantiation easy and modular: each stage of a pipeline or each role in a farm can be implemented as a separate function pod, with the skeleton’s logic connecting them. While this work focuses on Farm computation steering, we aim at a larger set of skeletons and have designed the function pods to be generic, reusable, and easily configurable to form other patterns. 

Our farm skeleton %(master-worker pattern), 
employs a template (see Sections~\ref{sec:formulation} and \ref{sec:farm:implem}) based on three types of functions: the Emitter, the Worker, and the Collector. Each one is a standalone OpenFaaS function (written in Python for our prototype) that knows how to interface with the others. As we detail in Section \ref{sec:farm:implem}, an initial setback was that the invocation mechanisms of the open-source version of OpenFaaS do not provide true scalability or efficient load balancing.
Moreover, FaaS advocates for the separation of computation from storage. Consequently, storage access and inter-function communication are achieved via \textit{mediated} channels, often through cloud storage or other indirect mechanisms~\cite{copik_fmi_2023}. Yet, efficient composition of serverless functions in general is complex and, without adding custom runtime support, it breaks the \textit{Serverless Trilemma}~\cite{baldini2017serverless}.
Since OpenFaaS functions are stateless and isolated, we decided to exploit its call semantics only to \emph{activate} pods that would then deal with passing data (tasks and results) between skeleton stages via a separate mechanism.

%Message queues

For this purpose we adopt the same Redis\footnote{See \url{https://redis.io/docs/latest/}} distributed queue service also exploited by OpenFaaS as the backbone connecting the function pods, leveraging its performance and proven compatibility. Once activated, each function reads from an input queue and pushes to an output queue, as specified by its skeleton-mandated configuration. This approach decouples producers and consumers, naturally supporting parallelism. Multiple worker instances can compete for tasks on the same queue (thus implementing the farm pattern), and a pipeline can be formed by linking one stage’s output queue to the next stage’s input queue. The use of a fast in-memory queue also enables asynchronous, streaming communication between functions, which is critical for farm and pipeline throughput.

By changing configuration (e.g. queue names, or environment flags), the same function pods can already be deployed and composed in different ways, enabling any combination of farms and pipelines while keeping the core logic the same. Additional pod designs can be easily coded to support other skeletons and templates.

\section{Problem formulation}\label{sec:formulation}
\begin{comment}
\noindent\todo{define the farm skeleton we address, its constraints and the parameters}
\todo{define our resource management (i.e., scaling ) problem}
\todo{Define the management problems. includes defining the metrics}
\todo{and defining how we quantify the objectives (performance, efficiency, QoS definitions and expression in terms of the metrics}\\

%%%%%%%%%%%
{\em
\begin{quote} 
    QoS Model (Deadline-Based) :
    Define a deadline model where the acceptable task completion time is calculated as:

        - Deadline = Coefficient (2 or 3) × Expected Duration × User Priority

    This allows adaptive deadlines based on task complexity and user-defined priority levels.
\end{quote}
}

\todo{what kind of baseline -- behavioural model of performance}

%%%%%%%%%%%
\todo{What kind of AI techniques we plan to use (RL ?)}
\todo{what features we do provide the AI (fairly at an abstract level, not all details are necessarily here).}    
\end{comment}

We consider a single instance of the stream-based \emph{farm skeleton} computational pattern ~\cite{aldinucci2017guide}, to be deployed on the OpenFaaS serverless platform.
%
%\subsection{Farm Skeleton Model}
%We adopt the \emph{farm skeleton} computational pattern%~\cite{aldinucci2012fastflow}
%~\cite{danelutto2014structured}, a well-established structured parallel programming paradigm for 
%stream-based applications, 
We adopt the same classic template as the cited FastFlow implementation %~\cite{aldinucci2017guide} 
%In our formulation, the farm 
comprising an input stage, a pool of parallel workers, and an output stage. In this work we focus on the worker pool size as the sole control parameter for dynamic resource management, while keeping all other configuration options fixed.
%
%\todo{instantiated as described in Section~\ref{sec:farm:implem}.}
%%
%, while the Emitter and Collector remain fixed.
%The farm skeleton, as implemented in the FastFlow framework~\cite{aldinucci2017guide}, consists of three primary components:
%
% \begin{itemize}
%     \item \textbf{Emitter} ($E$): Distributes incoming tasks from the input stream to a pool of parallel workers according to a load-balancing policy.
%     \item \textbf{Workers} ($W_1, W_2, \ldots, W_n$): A dynamic set of $n$ identical replicas that process tasks independently in parallel.
%     \item \textbf{Collector} ($C$): Aggregates results from workers and produces the output stream.
% \end{itemize}
%
%
In our template the worker pool size is constrained to an interval $N_{\min} \leq N_k \leq N_{\max}$ and can only be adjusted at discrete control steps separated by a fixed step duration $T_{step}$. At each step $k$, the controller issues a scaling action that changes the worker count by a small integer increment (scale down, no-op, scale up). In practice, the OpenFaaS deployment introduces non-negligible scale-up delay due to function and queue-worker initialization (see Section~\ref{sec:farm:implem}), so the effect of large scaling actions on the effective worker pool may be smoothed over multiple steps.

%\todo{add link to implementation section? ~\ref{sec:farm:implem}}
%
%\todo{Basic Perf. model}

Generic performance models of the farm skeleton are well studied and provide the foundation for dynamically steering resource allocation by selecting the next action according to the skeleton status. We gather a set of metrics that allow both reactive and RL-based policies to be evaluated and trained, capturing the internal state, resource allocation, and performance behaviour.

%the length of its internal queues lengths, % (input, worker, result, and output queues), the current number of Workers, summary statistics of processing times, arrival-rate estimates, and QoS indicators.
%resource allocation and 
At each control step $k$ we summarize the farm state as

\begin{equation}
\begin{aligned}
\mathbf{s}_k ={}
\big[\,&|Q_{in,k}|, |Q_{work,k}|, |Q_{res,k}|, |Q_{out,k}|, N_k,\\&\bar{T}_{proc,k}, T_{proc,k}^{max}, \lambda_{a,k}, q_k\,\big]^\top
\in \mathbb{R}^9.
\label{eq:state}
\end{aligned}
\end{equation}

Here all symbols with subscript $k$ denote quantities measured at control step $k$: $|Q_{\ldots,k}|$ are queue lengths, $\bar{T}_{proc,k}$ is the moving average of recent processing times, $T_{proc,k}^{max}$ is the maximum observed processing time in a sliding window, $\lambda_{a,k}$ is the moving average arrival rate, and $q_k$ is the instantaneous QoS defined as the fraction of tasks meeting their service time deadline within the step.
From these, we can derive three broad performance objectives:
\begin{itemize}
    \item \textbf{QoS}: ensure a set fraction of tasks complete before their deadlines, keeping the QoS above a target level $q^*$.
    \item \textbf{Responsiveness}: limit work backlog and latency by reacting to queue build-up and related waiting times.
    \item \textbf{Efficiency}: minimize the average %number 
    of active Workers and avoid excessive scaling actions while meeting QoS.
\end{itemize}

The dynamic scaling problem can thus be cast as a sequential decision problem: at each control step, the autoscaler observes the current farm state and chooses a scaling action according to a given policy that balances the tradeoff between QoS, responsiveness, and efficiency over time. 

%In the rest of this section we detail the farm skeleton model, the workload and QoS model, and the Markov Decision Process (MDP) formulation used for our reinforcement-learning (RL) controllers.

%%%%%%%%%%%%%%%%%%% use case begin
%\subsection{Use-case Modeling}\label{sec:usecase}
\smallskip
The farm workload in our experiments is a stream of image processing tasks that are described in Section~\ref{sec:experimental-settings}.
%
%
%The farm processes a continuous stream of image-processing tasks and consists of an Emitter, a pool of homogeneous Workers, and a Collector. The Emitter injects tasks into the system according to a time-varying arrival process; the Workers execute the calibrated image-processing pipeline; the Collector assembles results and evaluates their quality-of-service (QoS) status. The main control variable is the number of active Workers $n(t)$ over time; the Emitter and Collector are kept fixed.
%
Each task $\tau_i$ arrives at time $t_i^{arr}$ with an associated input size $s_i$ (length in pixels of the side of a square image) and an expected sequential processing time $\hat{T}_s(s_i)$.% obtained from the calibration model of Section~\ref{sec:experimental-settings}. 
We set a deadline $D_i$ for each task as a multiple of the expected duration,
\begin{equation}
    D_i = \beta \cdot \hat{T}_s(s_i)
\end{equation}
where $\beta > 1$ is a slack coefficient (we set $\beta=2$ in all experiments). % and $\text{prio}_i$ encodes the task's priority.
A task is said to meet its QoS requirement if its completion time $C_i$ satisfies $C_i - t_i^{arr} \leq D_i$. Over a control interval, we track the fraction of tasks that meet their deadlines as the instantaneous QoS level $q_k$ at control step $k$.

%%%%%%%%%%%%%%%%% use case end

%\subsection{Workload Model}
%Each task $\tau_i$ is characterized by:
%\begin{itemize}
%    \item \textbf{Arrival time}: $t_i^{arr}$ (when task enters $Q_{work}$)
%    \item \textbf{Data size}: $s_i$ (in pixels, for image processing workload)
%    \item \textbf{Deadline}: $D_i$ (maximum allowed completion time)
%\end{itemize}
\smallskip
%\subsection{Reinforcement Learning Formulation}
We model the scaling problem as a Markov Decision Process (MDP) described as a tuple $\langle \mathcal{S}, \mathcal{A}, \mathcal{P}, \mathcal{R}, \gamma \rangle$~\cite{sutton2018reinforcement}, where the elements are respectively state space, action space, transition function, reward function, and discount factor. The concrete realization of this MDP as a Gymnasium environment is described in Section~\ref{subsec:env}.
The objective is to find the optimal policy $\pi^*$ that maximizes the cumulative reward:
\begin{equation}\label{eq:rl_objective}
    \pi^* = \arg max_{\pi}\mathbb{E}\left[\sum_{k=1} \mathcal{R}^{\pi}_k\right]
\end{equation}
where $\mathcal{R}^{\pi}_k$ is the reward obtained following policy $\pi$ at control step $k$. 
To match the performance objectives with Eq. \ref{eq:rl_objective}, we designed the reward function accordingly (Eq.~\ref{eq:reward} in Fig.~\ref{fig:eq:reward}).

\section{Farm Skeleton Implementation}\label{sec:farm:implem}
%Assigned Constraints (e.g. composition capability)\\  %not for the farm case
% \todo{Design Choices and our reference farm template\\
% Function/process structures\\
% implementation choices}

\begin{comment}
The design and implementation of a Farm parallel skeleton, comprising Emitter, Workers, and Collector components, are built on top of OpenFaaS to handle image-analysis workloads under dynamic load conditions. Following the structured parallelism paradigm of the FastFlow\cite{FastFlow} Farm template, the system integrates a customized invocation and scaling workflow to achieve deterministic pod utilization and resilient autoscaling.

We adopt the Farm template as formalized in FastFlow, where an Emitter produces tasks, a pool of Workers processes tasks in parallel, and a Collector aggregates results. The template separates concerns cleanly: generation, parallel processing, and aggregation. We retain this division and use Redis-backed queues to instantiate the template’s bounded buffers between stages. In practice, we found that a plain “one-function-name with n replicas” approach on OpenFaaS did not guarantee fair utilization across pods. Thus, we refactored invocation and scaling to preserve the Farm’s predictability and efficiency under serverless deployment.    
\end{comment}

The farm skeleton is implemented on OpenFaaS following the FastFlow\cite{aldinucci2017guide} Farm template, comprising Emitter, Workers, and Collector components that handle image-analysis workloads under dynamic load conditions. The template separates generation, parallel processing, and aggregation; we retain this structure and connect stages via Redis-backed queues. An initial ``one-function-name with multiple replicas'' mapping in OpenFaaS led to unbalanced pod utilization, motivating the customized invocation and scaling workflow described in the next subsection.
%\todo{let's settle which one is the reference to farm in fastflow; we have 3, 4, and 32. 32% and 4 look redundant (same authors, almost same title). }
% We retain \cite{aldinucci2017guide} solely for the recommended FastFlow citation, as indicated in the official repository (https://github.com/fastflow/fastflow).

\begin{figure}[b]
    \centering
    \includegraphics[width=1.0\linewidth]{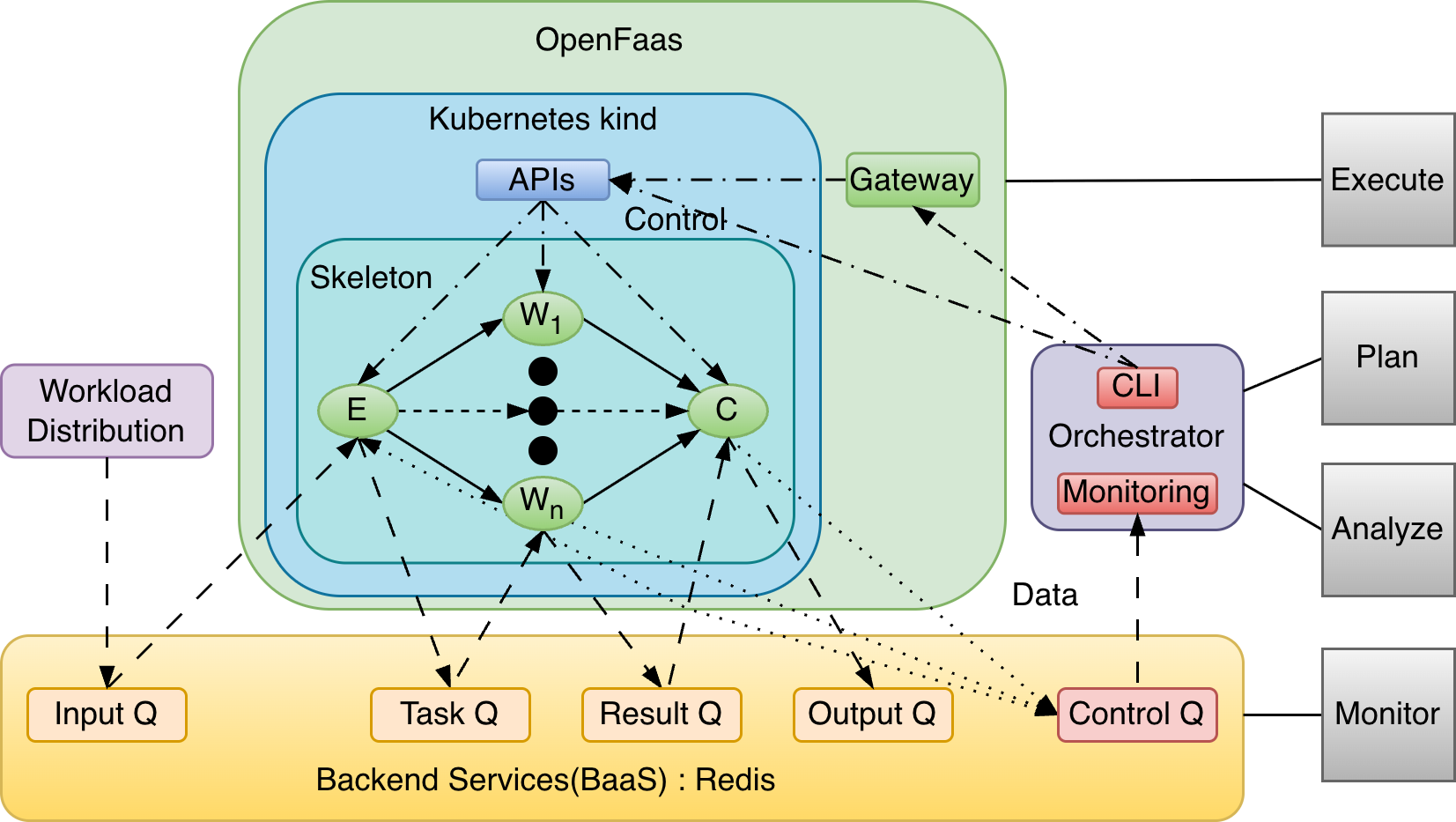}
    \caption{OpenFaaS-based farm skeleton template.}
    \label{fig:skleton}
\end{figure}

\smallskip
%\subsection{System Architecture}
The farm implementation consists of three OpenFaaS functions and an external autoscaling environment (orchestrator), as illustrated in Fig.~\ref{fig:skleton}.

\begin{comment}  % described twice in a page 
\begin{itemize}

  \item \textbf{Emitter} ($E$) produces tasks per workload phase and enqueues them into \texttt{worker\_queue}.
  \item \textbf{Workers} ($W_1, W_2, \ldots, W_N$) constitute a pool of single-replica function services, each pulling from \texttt{worker\_queue}, processing tasks, and pushing results to \texttt{result\_queue}.
  \item \textbf{Collector} ($C$) dequeues results, evaluates their QoS against per-task deadlines, and publishes annotated outputs to \texttt{output\_queue}.
  \item \textbf{Environment} orchestrates deployment, invocation, and per-step scaling actions for training and evaluation.

  \item \textbf{Emitter} ($E$) produces tasks and enqueues them into \texttt{worker\_queue}.
  \item \textbf{Workers} ($W_1, W_2, \ldots, W_n$) form a pool of functions that pull from \texttt{worker\_queue}, process tasks, and push results to \texttt{result\_queue}.
  \item \textbf{Collector} ($C$) dequeues results, checks QoS against per-task deadlines, and publishes outputs to \texttt{output\_queue}.
  \item \textbf{Environment} orchestrates deployment, invocation, and scaling actions during training and evaluation.

\end{itemize}
\end{comment}

The queues (\texttt{input\_queue} $\rightarrow$ \texttt{worker\_queue} $\rightarrow$ \texttt{result\_queue} $\rightarrow$ \texttt{output\_queue}) realize the bounded buffers between stages, providing natural back-pressure and decoupling producer/consumer rates. All functions use a shared \texttt{program\_start\_time} for consistent relative timestamps. In addition to these data queues, a dedicated set of control queues (e.g., \texttt{start-queue}) connects the environment with the Emitter, Workers, and the Collector for lifecycle and scaling commands (e.g., start, scale, terminate).

\subsection{Customized Invocation and Scaling}

% Moved to appendix
A straightforward strategy that invoked a single function name with multiple replicas and relied on OpenFaaS/Kubernetes Services for load balancing led to uneven utilization, with the gateway reusing HTTP keep-alive connections to a single pod and leaving others idle under load.

To obtain deterministic load distribution and explicit control over scaling, we instead deploy $N$ 
distinct worker functions (\texttt{worker-1}, \dots, \texttt{worker-n}), each with a single replica, and invoke them asynchronously via the OpenFaaS API under the control of the external environment. Each invocation maps one-to-one onto a pod and launches a long-running handler loop, potentially keeping the worker function alive indefinitely. The environment manages scale-up and scale-down as one-step changes in the active worker set while preserving in-flight work.

%The observed bottlenecks are rooted in the scale-up path of our OpenFaaS deployment. 
The asynchronous call strategy that allows managed scaling in our design requires a one-to-one pairing between each Worker function and a dedicated \texttt{queue-worker} replica that drives its asynchronous invocation throughout the Worker’s lifecycle. 
The dedicated queue manager is necessary to decouple function execution from HTTP connection lifetimes, and prevent opportunistic termination or scale-to-zero of long-running handlers by OpenFaaS. 
%preserve a stable asynchronous delivery path, 

In practice, end-to-end scale-up for a single Worker takes about 5 to 8 seconds (covering Kubernetes scheduling, function cold start, queue-worker startup, and completing the handshake), but 
%the  start-queue acknowledgment that confirms the Worker has entered its long-running loop. 
%When multiple Workers are added, 
the overall time grows approximately linearly with the number of Workers. 
%
%Thus, scaling up by $x$ Workers also entails scaling up by $x$ queue workers and completing related handshakes. 
%
The trade-off of efficient task distribution is a near-linear scale-up latency that can challenge timely reactions to high load.
%
%because each new replica requires its own queue-worker and acknowledgment. 
Under abrupt load spikes, the orchestrator may therefore lag in raising capacity quickly enough, leading to transient queue growth and an elevated risk of missed deadlines.

\subsection{Function and Process Realization}

\begin{comment}
\textbf{Emitter (producer).} The Emitter ingests phase descriptors from \texttt{input\_queue}, each defining a target number of tasks, arrival characteristics, and deadlines. For each phase, it generates tasks using controlled stochastic timing and pushes them into \texttt{worker\_queue}. Reproducibility is achieved by seeding the random number generators when a \texttt{task\_seed} is provided. The Emitter runs as a long-lived loop until a control message instructs it to terminate.

\textbf{Workers (parallel stage).} Each Worker instance independently pops tasks from \texttt{worker\_queue}, executes the image-processing pipeline, records processing times, and enqueues results to \texttt{result\_queue}. The Worker monitors control channels for \texttt{SCALE\_DOWN/SYN} and responds with \texttt{ACK} when it reaches a safe point, then exits; it also honors \texttt{TERMINATE}. Readiness is signaled via a start-queue handshake that acknowledges completion of scale-up; it confirms that each newly added worker has entered its long-running loop and is ready to accept tasks.

\textbf{Collector (aggregator).} The Collector pops results from \texttt{result\_queue}, reconstructs end-to-end timing relative to \texttt{program\_start\_time}, and evaluates QoS by comparing each task's completion time with its deadline. It writes annotated outputs to \texttt{output\_queue} and logs QoS status. The Collector also participates in start and termination signaling to align the time origin and shut down cleanly.
\end{comment}

\textbf{Emitter.} The Emitter receives tasks from outside the skeleton and dispatches them to the Workers. In our simulation we also integrate task production into this component to avoid unnecessary Redis overhead. The Emitter pod reads per-phase descriptors from the \texttt{input\_queue}, generates tasks with controlled stochastic timing, and enqueues them into the \texttt{worker\_queue}. It runs as a long-lived loop and supports seeded runs for reproducibility.

\textbf{Workers} ($W_1, W_2, \ldots, W_N$) form a pool of functions that compute in parallel. Each Worker pulls tasks from the \texttt{worker\_queue}, executes the image-processing pipeline, records processing times, and sends results to \texttt{result\_queue}. Control signals from the environment govern when Workers are added, drained, or terminated, but the data path remains a simple pull--process--push loop.

\textbf{Collector.} The Collector consumes completed tasks from the \texttt{result\_queue} and dispatches them to the \texttt{output\_queue}. Additional farm semantics (e.g.~task reordering) are not needed in our experiments, but the Collector hosts monitoring code. In our prototype it reconstructs end-to-end timings relative to \texttt{program\_start\_time}, evaluates QoS against per-task deadlines, annotates the output, and logs QoS status. It also participates in synchronizations to align the time origin at skeleton startup and shutdown.

\textbf{Orchestrator.} The Orchestrator implements the autoscaling environment of Section~\ref{subsec:env} and hosts the autoscaler. It runs outside the OpenFaaS functions, periodically reading queue lengths, worker counts, timing summaries, and QoS indicators, and forwards them to an autoscaler implemented either as a reactive baseline (ReactiveAverage, ReactiveMaximum or as a reinforcement-learning agent (SARSA, DQN in Section~\ref{sec:ai_approaches}). The selected policy then issues scale-up/scale-down requests through the OpenFaaS API. This Gymnasium-style control layer keeps the farm data path separate from the management plane, as detailed in Section~\ref{subsec:env}.

\subsection{Workload and QoS Modeling}
The workload follows a four-phase arrival pattern introduced in Section~\ref{sec:experimental-settings}, combining phases with different steady and oscillatory regimes as summarized in Fig.~\ref{fig:experiment4phases} and Tab.~\ref{tab:arrival_pattern}. Each task is annotated with an expected processing time and a derived deadline from the calibrated image-processing model, and QoS decisions in the Collector compare completion times against these deadlines. Optional phase shuffling supports robustness experiments.

\subsection{Autoscaling Environment}
\label{subsec:env}
The environment provides a Gymnasium-style interface for experimentation with reinforcement learning and reactive baselines. The observation vector $\mathcal{S}$ defined by Eq.~\ref{eq:state} comprises queue depths, worker count, timing summaries, estimated arrival rates, and QoS indicators. Actions are single-step deltas in the worker pool size, clipped to configured bounds ($\mathcal{A} = \{-1, 0, +1\}$) and subject to step-duration constraints. The environment deploys and invokes functions, manages queue-worker capacity, and returns per-step information such as queue depths, current workers, processing-time summaries, estimated arrival rate, and QoS success rate as a 9-element vector. The reward function combines QoS tracking, backlog control, worker efficiency, and scaling friction with configurable weights and thresholds. This external control layer keeps scaling orthogonal to the farm's data plane, enabling reproducible studies under identical pipeline semantics.

\subsection{Implementation Considerations and Limitations of the Farm Skeleton on OpenFaaS}
We ensure resilient Redis operations for queue interactions, employing safe calls and limited retries to handle transient errors without violating pipeline invariants. Control channels are separated from data queues to simplify reasoning about liveness and safety. 
On scale-up, the environment coordinates worker-service deployment with adjustments to the OpenFaaS \texttt{queue-worker} so that asynchronous invocations are not delayed. All components timestamp events are relative to \texttt{program\_start\_time}, ensuring coherent end-to-end tracing across functions and enabling precise QoS evaluation.

\begin{figure}[t]%ht
    \centering
    \includegraphics[trim={0 9 0 7},clip,width=1\linewidth]{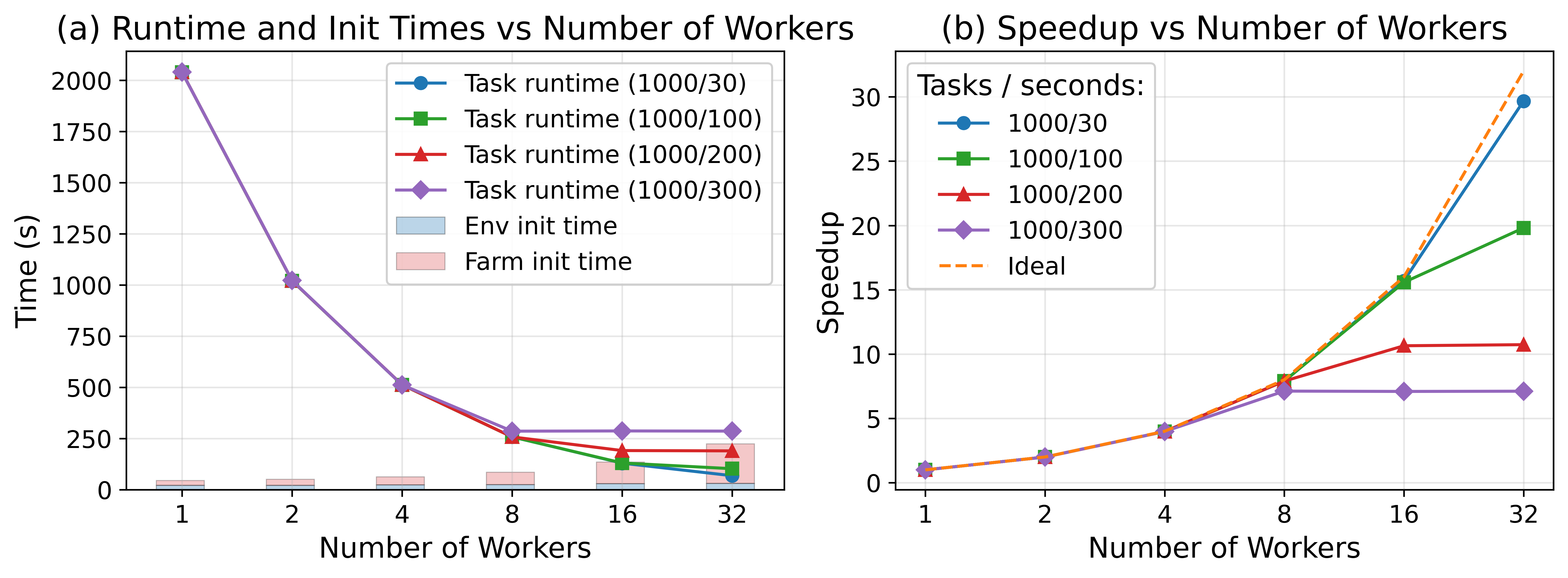}
    \caption{Scaling metrics for the farm skeleton: runtime, initialization overhead (left) and speedup (right) vs. worker count, under static workloads.}
    \label{fig:scaling_metrics}
\end{figure}

To validate the correctness and scalability of the farm skeleton implementation, we first run experiments with varying numbers of workers and workload configurations. Fig.~\ref{fig:scaling_metrics} summarizes runtime and initialization overheads, showing that while task runtime decreases with increasing parallelism, farm initialization time grows with the number of Workers. While in infinite-stream settings initialization can be ignored, the OpenFaaS-induced scale-up latency imposes a practical ceiling on effective parallelism for finite streams as well as dynamically varying workloads, motivating the need for autoscaling policies that trade off throughput gains against initialization costs.

\begin{comment}
\subsection{Limitation: Scale-up Latency and Orchestration Overhead}
The current design requires a one-to-one pairing between each Worker function and a dedicated \texttt{queue-worker} replica that drives its asynchronous invocation throughout the Worker’s life cycle. As a result, scaling up by k Workers also entails scaling up by k \texttt{queue-worker} replicas and completing a readiness handshake for each new Worker before it can accept tasks.

In practice, end-to-end scale-up for a single Worker takes about 5 to 8 seconds, covering Kubernetes scheduling, function cold start, \texttt{queue-worker} startup, and the start-queue acknowledgment that confirms the Worker has entered its long-running loop. When multiple Workers are added, the overall time grows approximately linearly with the number of Workers because each new replica requires its own \texttt{queue-worker} and acknowledgment. Under abrupt load spikes, the orchestrator may therefore lag in raising capacity quickly enough, leading to transient queue growth and an elevated risk of missed deadlines.

Despite this overhead, the dedicated queue-worker per Worker is necessary to preserve a stable asynchronous delivery path, decouple function execution from HTTP connection lifetimes, and prevent opportunistic termination or scale-to-zero of long-running handlers. The trade-off is a near-linear scale-up latency that can challenge timely reactions to high load.    
\end{comment}

\section{Farm Dynamic Management Policies}
% \todo{Baseline with reactive policy\\
% AI approach, what is SARSA, why we choose it, how it is implemented}

\begin{comment}
The dynamic management of the Farm skeleton contrasts reactive baselines with learning-based agents. The Farm scales its worker pool online based on queue dynamics, arrival rates, and QoS feedback, applying each scaling decision as a one-step adjustment to the worker count.
\end{comment}
We now define the autoscaling policies used in our experiments: analytical reactive baselines and learning-based agents.
%
%
%\subsection{Baseline: Reactive Policies}
We implemented two analytical baselines that compute the required parallelism directly from queue lengths and service-time estimates. Both policies consume the environment observation vector and output an action in \{-1, 0, +1\} via the environment’s discrete mapping.

\texttt{ReactiveAverage} (RA) uses the average processing time as the service-time estimate. It plans over a finite horizon and provisions capacity for both clearing the current backlog and serving expected arrivals. Let \texttt{service\_time} be the average processing time and \texttt{horizon} the planning window. It computes workers for arrivals as $arrival\_rate * service\_time$ and workers for backlog as $(worker\_q / horizon) * service\_time$. 
%The sum, multiplied by a safety factor, forms the target worker count. 
A mirrored enqueue counter provided by the environment is used to estimate active tasks and convert them into effective busy capacity, which refines the idleable capacity before mapping the difference to a discrete action.

%\[
%workers_for_arrival = arrival_rate * service_time
%
%\]

\texttt{ReactiveMaximum} (RM) mirrors the above logic but uses the maximum observed processing time in the state. It emphasizes QoS protection by provisioning against worst-case service time. 
%An adjustment factor and safety factor modulate sensitivity and conservativeness. 
Both policies maintain negligible computational overhead and require no training, making them useful baselines and operationally attractive when training is not possible.

\begin{figure}[t]
%    \centering
%
\begin{align}
R_k ={}&
\underbrace{w_{\text{qos}}(q_k - q^*)}_{\text{QoS tracking}}
-
\underbrace{w_{\text{backlog}}\!
\left(\frac{Q_k}{Q^*}-1\right)^2}_{\text{Queue/backlog penalty}}
-
\underbrace{w_{\text{scale}}}_{\text{Scaling cost}}
\notag \\
&-
\underbrace{
w_{\text{eff}}\,
\mathbf{1}\!\left[Q_k \le Q_{\text{idle}},\; q_k \ge q^*\right]
\max(0,\, N_k - N^*)
}_{\text{Over-provisioning penalty}}
\notag \\
&+
\underbrace{
w_{\text{up}}\, 
\mathbf{1}\!\left[\Delta N_k > 0,\; \left(\frac{Q_k}{Q^*} > 1 \;\text{or}\; q_k < q^*\right)\right]
}_{\text{Reward for justified scale-up}}
\notag \\
&+
\underbrace{
w_{\text{down}}\, 
\mathbf{1}\!\left[\Delta N_k < 0,\; \left(\frac{Q_k}{Q^*} \le 1 \;\text{and}\; q_k \ge q^*\right)\right]
}_{\text{Reward for safe scale-down}}
\notag \\
&+
\underbrace{
w_{\text{qos}}\,
\mathbf{1}\!\left[q_k \ge q^*,\, \frac{Q_k}{Q^*}\le 0.5\right]
}_{\text{Bonus for stable high-QoS operation}}.
\label{eq:reward}
\end{align}
\hrulefill\par
\caption{The reward function for SARSA and DQN}\label{fig:eq:reward}
\end{figure}

To clarify the impact of reward shaping, we note that the main weights in Eq.~\eqref{eq:reward} act as intuitive knobs that shift the QoS--cost--stability trade-off. Increasing $w_{\text{qos}}$ and $w_{\text{backlog}}$ makes the controller more aggressive in matching deadlines and suppressing queue growth, typically at the cost of higher mean workers and more frequent scale-ups. Increasing $w_{\text{scale}}$ discourages reconfigurations and can improve stability, but if set too high it delays corrective actions and may reduce QoS under fast load changes. The efficiency term $w_{\text{eff}}$ mainly affects low-load behaviour by encouraging scale-down once QoS is satisfied and $Q_k$ is small, while the directional incentives ($w_{\text{up}}$, $w_{\text{down}}$) bias the policy toward earlier scale-up or more conservative scale-down. In our evaluation we keep the reward coefficients fixed to enable controlled comparisons across policies.

\subsection{AI Approach: SARSA and DQN}
\label{sec:ai_approaches}
We instantiate two standard value-based RL agents for scaling: a tabular SARSA($\lambda$) controller over a discretized state space and a lightweight Double DQN operating on the continuous observation vector, following established formulations for SARSA and deep Q-learning. Both interact with the same Gymnasium-style environment~\ref{subsec:env}, consume the state vector $\mathbf{s}_k$ defined in Eq.~\ref{eq:state}, and act on the same discrete scaling actions $\mathcal{A} = \{-1, 0, +1\}$. SARSA uses a discretized state representation with eligibility traces, whereas Double DQN uses a small MLP over continuous observations; in both cases we keep hyperparameters fixed across all evaluation runs.

To connect these autoscaling objectives to a learnable signal, we define a shaped reward that combines four primary components: QoS tracking relative to a target rate, queue backlog relative to a healthy target, worker efficiency relative to a balanced target, and a penalty on scaling events.

Equation \eqref{eq:reward} in Fig.~\ref{fig:eq:reward} defines a shaped reward that balances service quality, backlog control, resource efficiency, and stability.
The terms jointly reward high QoS and justified scale-up/scale-down decisions while penalizing backlog overflow, over-provisioning, and unnecessary scaling, guiding the agent toward timely but non-oscillatory scaling behavior.

\section{Experimental settings}
\label{sec:experimental-settings}

Our use case is an image-processing application with variable image sizes, implemented as a four-stage sequential pipeline (thumbnail generation, compression, metadata extraction, and format conversion). This provides a concrete workload whose service time scales with image size and underpins our QoS-deadline model.

To obtain hardware-agnostic service times, we profile the full pipeline on a reference host for the relevant image sizes and fit a reduced quadratic model from image size to end-to-end time. Table~\ref{tab:expected_time} reports the resulting expected times and default QoS deadlines used in the experiments.

\begin{table}[b]
    \centering
\caption{Example of expected model outputs $T_s$ and QoS deadlines}
\label{tab:expected_time}
    \begin{tabular}{lcccc}\toprule
%         Image Size& 512×512& 1024×1024&  2048×2048& 4096×4096 \\
         \textbf{Image Size}& 512×512& 1K×1K&  2K×2K& 4K×4K \\
          \midrule
         \textbf{Expected time}&  0.046 s& 0.181 s& 0.719 s& 2.870 s \\
         \textbf{Default deadline}\ ($\beta = 2$) &  0.09 s & 0.35 s &  1.42 s & 5.58 s\\ 
          \bottomrule
    \end{tabular}
\end{table}

% Moved to appendix
\begin{comment}
Our simulation uses these calibrated durations inside each Worker handler, where processing time is emulated with \texttt{time.sleep(processing\_time)}.
%
%the data plane without tying performance to the physical CPU core count. 
%
%In the Worker handler, each task’s calibrated processing time is emulated with \texttt{time.sleep(processing\_time)}. 
%
This simulation-based approach allows testing with virtually unlimited parallel Workers, bypassing bottlenecks from physical CPU core count and cluster size as long as the communication and platform overheads remain negligible. 
%limited cores on a development host or small cluster.
It preserves realistic service times and variance while isolating the effect of the dynamic management policy itself.
\end{comment}

In the farm simulation, each Worker emulates processing via \texttt{time.sleep(processing\_time)}, which allows us to isolate the impact of the autoscaling policy and OpenFaaS control-plane latencies. This abstracts away compute contention (e.g., CPU saturation and cache/memory effects) and noisy network/storage conditions, so the results primarily characterise the controller’s ability to cope with OpenFaaS control-plane latencies and queueing dynamics rather than compute-bound interference.

%  \todo{
%  Task Arrival Patterns
% The experiment will use four consecutive time windows, each representing a different task arrival pattern to simulate 
% dynamic workloads:\\
% \todo{maybe a table}
%     Phase 1 – Steady Low Load:
%         - Uniformly low task arrival rate.
%         - Represents an underloaded or idle system state.
%     Phase 2 – Steady High Load:
%         - Uniformly high task arrival rate.
%         - Represents a sustained heavy workload scenario.
%     Phase 3 – Slow Oscillation:
%         - Gradual increases and decreases in task arrival rate (low-frequency oscillations).
%         - Simulates workloads with smooth, periodic fluctuations in demand.
%     Phase 4 – Fast Oscillation:
%         - Rapid increases and decreases in task arrival rate (high-frequency oscillations).
%         - Represents a highly dynamic environment with frequent, short-term bursts of traffic.}

\subsection{Task Arrival Patterns and Sizes}

% Moved to appendix
\begin{comment}
The workload process is designed to stress control rather than compute. We keep the random distribution of per-task computational load fixed across experiments so that the difficulty of individual tasks is stationary. At the same time, we vary the number of tasks per second by programmatically modulating interarrival times according to four patterns shown in Fig.~\ref{fig:experiment4phases}. The goal is to stress the scaling policy under steady, ramping, and bursty regimes while holding the per‑task processing distribution fixed. This isolates the effect of the orchestrator from changes in task complexity. Each phase lasts a configurable duration, and within a phase the task generator uses controlled Poisson windows to modulate interarrival times and still hit the exact target task count. The base arrival rate and phase multipliers determine the aggregate load per phase; the generator computes per‑window expected counts and adjusts the last window to meet the target exactly. 
\end{comment}

\begin{figure}[t]
  \centering
  \includegraphics[trim={0 73mm 0 0 },clip,width=1\linewidth]{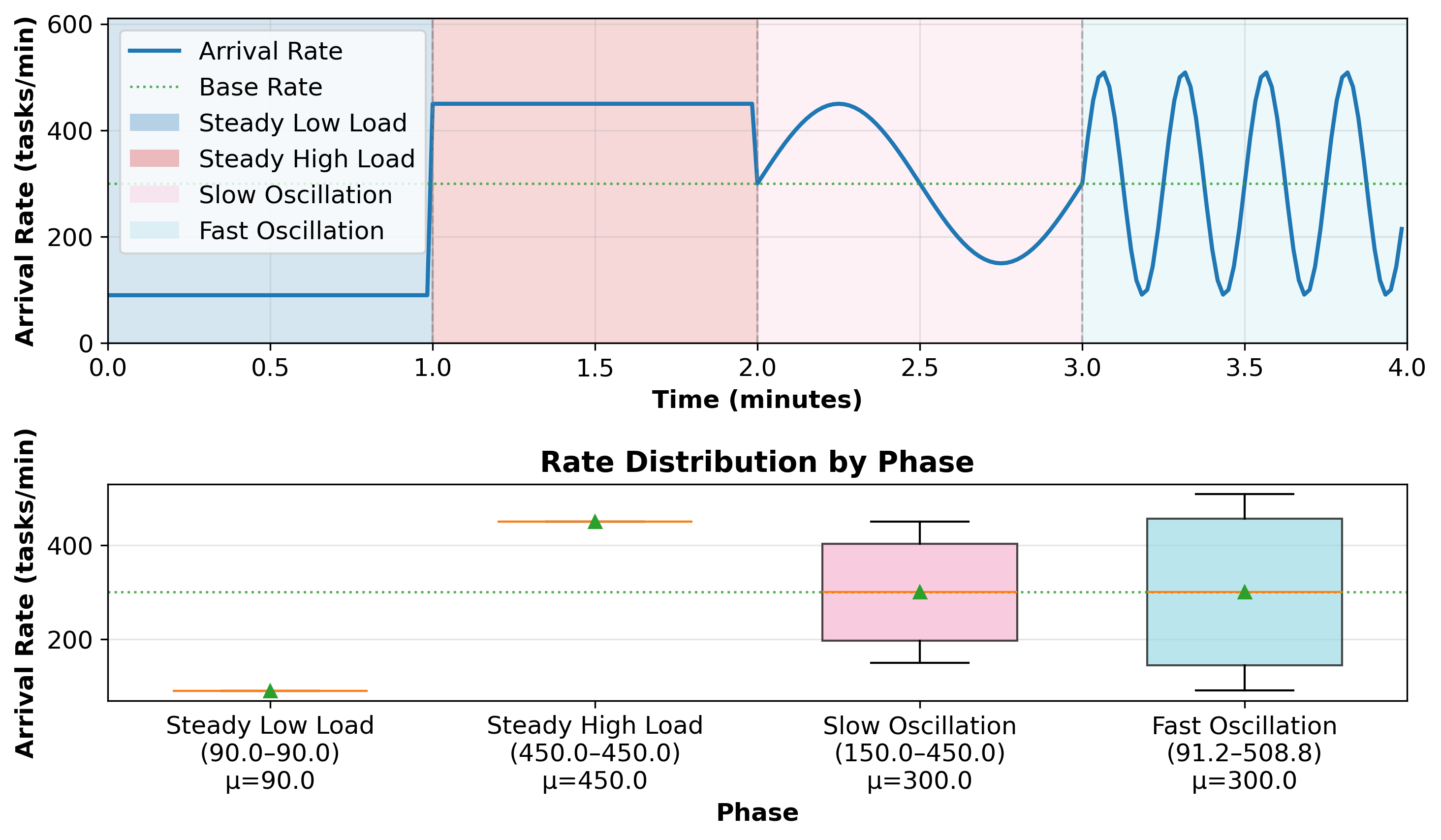}
  \caption{Configured mean of task arrival patterns across four workload phases.}
  \label{fig:experiment4phases}
\end{figure}

The workload process is designed to stress control rather than compute. We use a weighted random distribution of per-task work size, so that the mean task computational load is fixed to $1,5s$ (a stochastic stationary behaviour). We modulate the arrival rate across four phases (steady low, steady high, slow oscillation, fast oscillation) as summarized in Fig.~\ref{fig:experiment4phases} and Table~\ref{tab:arrival_pattern}. Within each phase we thus generate stochastic arrivals as a Poisson process (exponentially distributed inter-arrival times) with phase-appropriate rate, and enforce an exact per-phase task count via short windows with the final window adjusted to match the target total.

\begin{comment}
    \begin{tabular}{p{1cm}p{2cm}p{1cm}p{1cm}p{1cm}p{2.5cm}}%{|c|c|c|c|c|}\hline
%         Phase&  Arrival pattern&  Load level relative to base&  Oscillation cycles& Typical intent\\\hline
\toprule
         \textbf{Phase}&  \textbf{Arrival pattern}&  \textbf{Load level}  & \textbf{Avg. Tasks} &  \textbf{Cycles} & \textbf{Typical intent}\\%\hline
          &   &  \textbf{to base rate} & \textbf{/\,Minute} &  \textbf{/\,Phase} &  \\%\hline
          \midrule
         Steady Low Load&  Uniform interarrival times &  0.3 & 90 &  0& 	Underloaded or idle regime for stability checks\\\midrule%\hline
         Steady High Load&  Uniform interarrival times &  1.5 & 450 &  0& Sustained heavy load to test scale‑up capacity\\\midrule %\hline
         Slow Oscillation&  Sinusoidal modulation, low frequency&  0.5–1.5 & 300 &  1  & Smooth, periodic fluctuations to test tracking\\\midrule %\hline
         Fast Oscillation&  Sinusoidal modulation, high frequency&  0.3–1.7 & 300 &  4  & Rapid bursts to test responsiveness and damping\\ %\hline
         \bottomrule
    \end{tabular}
\end{comment}

\begin{table}[b]%[b]
    \centering
\caption{Task arrival patterns in the four phases of each episode.}
\label{tab:arrival_pattern}
\scriptsize
    \begin{tabular}{p{2cm}@{ }c@{ }c@{ }c@{ }p{28mm}}%{|c|c|c|c|c|}\hline
%         Phase&  Arrival pattern&  Load level relative to base&  Oscillation cycles& Testing purpose\\\hline
\toprule
         \textbf{\hfil Phase,\newline Arrival pattern}&  \parbox[t]{11mm}{\centering\textbf{Load wrt base level}}  & \parbox[t]{12mm}{\centering\textbf{Avg. Tasks /min.}} &  \parbox[t]{1cm}{\centering\textbf{Cycles\newline/Phase}} & \hfil \textbf{Typical intent}\\%\hline
%            &  \textbf{to base rate} & \textbf{/\,Minute} &  \textbf{/\,Phase} &  \\%\hline
          \midrule
         0:\,\,Steady Low Load, Poisson &  0.3 & 90 &  -& 	Idle regime, for stability checks\\\midrule%\hline
         1:\,\,Steady\,High\,Load, Poisson &  1.5 & 450 &  -& Sustained heavy load, test scale‑up capacity\\\midrule %\hline
         2:\,\,Slow Oscillation, Sinusoidal Poisson&  0.5–1.5 & 300 &  1  & Smooth  fluctuations, test tracking\\\midrule %\hline
         3:\,\,Fast Oscillation, Sinusoidal Poisson&  0.3–1.7 & 300 &  4  & Rapid bursts, test responsiveness, damping\\ %\hline
         \bottomrule
    \end{tabular}
        
\end{table}

\subsection{Evaluation Methodology}
Each policy was evaluated over 10 independent episodes (240 seconds each) with different random seeds. 
%
%
%We report mean $\pm$ standard deviation over the 10 runs for QoS success rate (fraction of tasks meeting deadlines), resource usage (mean and maximum worker count), and stability (scaling actions and no-ops); 
% PICTURE IS GONE! %for learning-based approaches we also report cumulative reward.
%\subsection{Experimental Results}
Table~\ref{tab:comparison} summarises metrics over the episodes (mean $\pm$ standard deviation) for the four scaling policies, reporting QoS success rate (fraction of tasks meeting deadlines), resource usage, and stability (scaling actions and no-ops).

The learning‑based agents clearly outperform the reactive baselines in final QoS: SARSA reaches $99.20\% \pm 0.52\%$ and DQN $94.91\% \pm 3.79\%$, whereas RA and RM achieve only $50.87\% \pm 4.38\%$ and $65.61\% \pm 14.26\%$, respectively. SARSA therefore offers the strongest guarantees, and DQN still improves QoS by more than 40 percentage points over both reactive policies. This strong SARSA performance, however, relies on a carefully designed discretisation of the observation space, whereas DQN operates directly on the continuous observation vector, trading a small QoS loss for a simpler interface between environment and controller.

\begin{table}[b]
\centering
\caption[Comparison of farm scaling policies with reactive baselines.]{Comparison of farm scaling policies with reactive baselines overall and as per-phase aggregates
%over the four arrival-pattern phases
%(Table~\ref{tab:arrival_pattern}) 
of QoS and mean workers. All metrics reported as mean $\pm$ STD over 10 runs.}
%{ and computed from the step-level evaluation logs.}
\label{tab:comparison}
\label{tab:per_phase}
\scriptsize
\begingroup
\setlength{\tabcolsep}{3pt}
%\resizebox{\columnwidth}{!}
{
\begin{tabular}{@{}c@{}l@{\,\,}c@{\,\,\,\,}c@{\,\,\,\,}c@{\,\,}c@{}}
\toprule
\textbf{Phase} & \hfil\textbf {Metric} & \textbf{SARSA} & \textbf{DQN} & \textbf{ReactiveAvg} & \textbf{ReactiveMax} \\
\midrule
&Final QoS [\%]      & $99.20 \pm 0.52$  & $94.91 \pm 3.79$& $50.87 \pm 4.38$ & $\,\,65.61 \pm 14.26$ \\
&Mean Workers        & $14.78 \pm 0.49$  & $14.40 \pm 0.77$& $10.79 \pm 0.31$ & $13.92 \pm 0.68$ \\
&Max Workers         & $17.80 \pm 0.75$  & $15.60 \pm 1.02$& $16.60 \pm 0.66$ & $18.80 \pm 0.40$ \\
&Scaling Actions     & $24.00 \pm 2.05$  & $\,\,3.60 \pm 1.02$& $28.40 \pm 0.92$ & $26.80 \pm 1.25$ \\
&No-op Actions       & $\,\,4.90 \pm 1.92$   & $24.60 \pm 1.69$& $\,\,1.00 \pm 1.00$  & $\,\,1.70 \pm 0.78$ \\
\midrule
\multirow{2}*{0} & QoS [\%] & $99.7 \pm 0.5$ & $99.2 \pm 1.1$ & $91.3 \pm 4.7\ $ & $91.0 \pm 5.1\ $ \\
  & mean workers & $14.26 \pm 0.31$ & $12.01 \pm 0.04$ & $8.11 \pm 0.14$ & $8.91 \pm 0.59$ \\
\multirow{2}*{1}  & QoS [\%] & $99.0 \pm 0.8$ & $90.1 \pm 7.5$ & $2.0 \pm 2.9$ & $27.0 \pm 32.4$ \\
%Steady High 
& mean workers & $16.66 \pm 0.35$ & $14.54 \pm 1.12$ & $10.30 \pm 0.98$ & $12.88 \pm 1.63\ $ \\
\multirow{2}*{2}  & QoS [\%] & $99.8 \pm 0.2$ & $99.9 \pm 0.2$ & $76.6 \pm 17.1$ & $95.8 \pm 10.6$ \\
%Slow Oscillation 
& mean workers & $13.36 \pm 0.90$ & $15.60 \pm 1.02$ & $14.44 \pm 1.30$ & $17.27 \pm 0.46$ \\
\multirow{2}*{3}  & QoS [\%] & $99.7 \pm 0.3$ & $99.9 \pm 0.3$ & $93.0 \pm 4.6$ & $100.0 \pm 0.1\ $ \\
%Fast Oscillation 
& mean workers & $14.63 \pm 1.20$ & $15.60 \pm 1.02$ & $10.39 \pm 0.46$ & $16.96 \pm 0.89$ \\
\bottomrule
\end{tabular}
}
\endgroup
\end{table}

Table~\ref{tab:per_phase} also helps pinpoint where each policy succeeds or fails under the four arrival regimes. SARSA maintains near-perfect QoS in every phase while keeping mean workers in the mid-teens, indicating that it scales up promptly in the sustained high-load phase and scales down again during oscillatory phases without accumulating backlog. DQN attains similarly high QoS in three phases but drops during sustained high load, consistent with its conservative scaling behaviour (few reconfigurations) and the non-negligible OpenFaaS scale-up latency, which together can delay capacity increases when the load remains high. The reactive baselines show their largest degradation in the steady high-load phase: despite scaling activity, they under-provision relative to the sustained arrival rate, leading to large queues and missed deadlines; their partial recovery in oscillatory phases reflects that the load periodically returns to more manageable levels.

\begin{figure}[t]
    \centering
    \begin{minipage}[t]{0.47\textwidth}
        \centering
        \includegraphics[trim={0 0 0 7mm},clip,width=\linewidth]{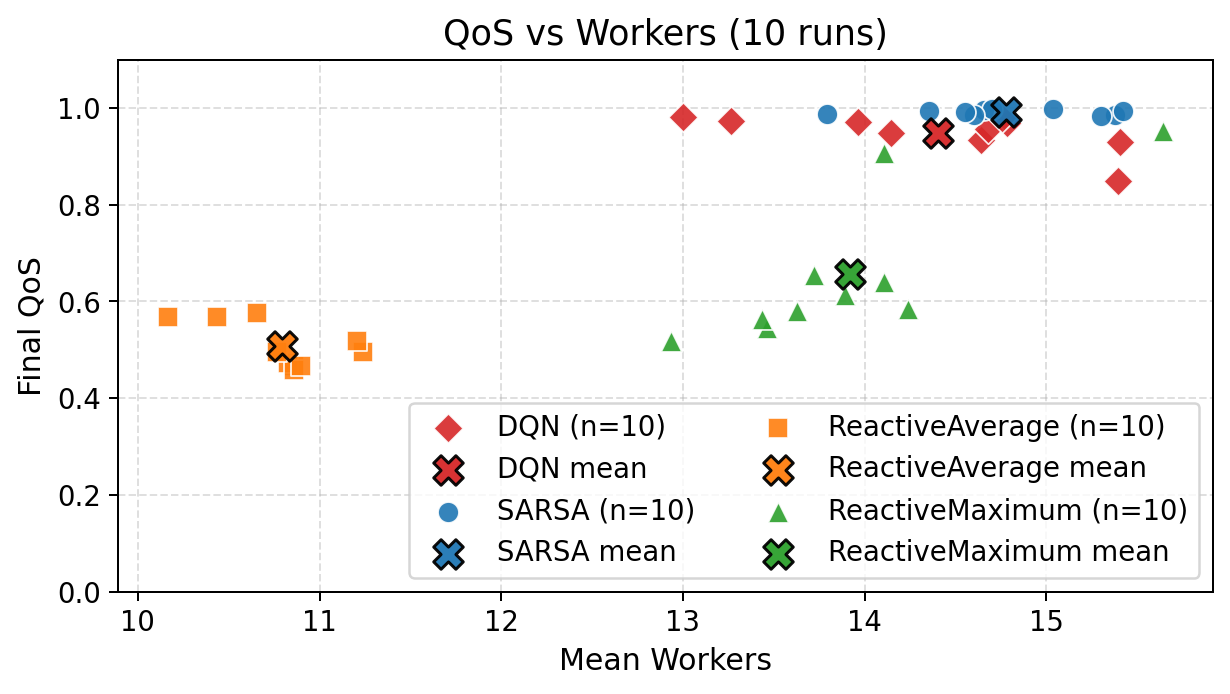}
    \end{minipage}\\%\hfill
    \begin{minipage}[t]{0.47\textwidth}
        \centering
        \includegraphics[trim={0 0 0 8mm},clip,width=\linewidth]{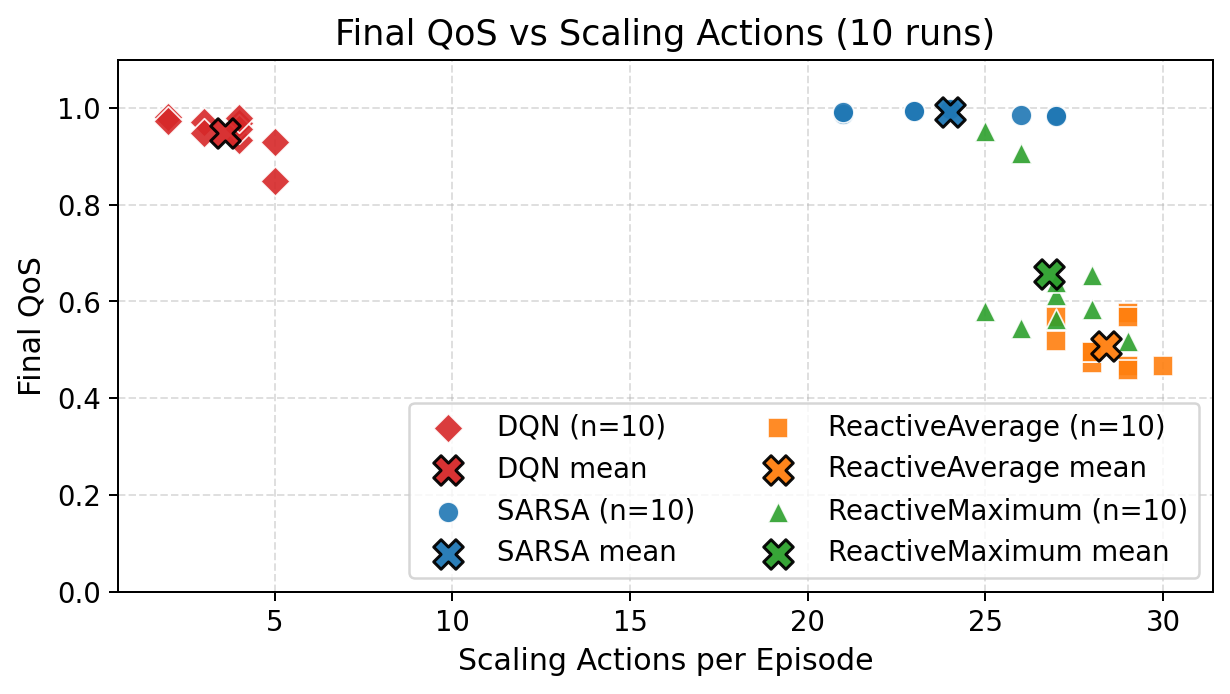}
    \end{minipage}
    \caption{Final QoS vs. mean workers (up) and number of scaling actions (down), over 10 runs (per-run points and overall average).}
    \label{fig:qos_tradeoffs}
\end{figure}

The QoS–resource trade‑off in Fig.~\ref{fig:qos_tradeoffs} (up) shows that SARSA combines the highest QoS with moderate average workers, whereas DQN uses similar capacity but settles at a slightly lower QoS. The reactive baselines either under-provision (RA) or spend comparable capacity without matching the QoS of the learning-based policies (RM).
Fig.~\ref{fig:qos_tradeoffs} (down) highlights that SARSA reaches near-optimal QoS with fewer scaling actions than the reactive baselines, while DQN achieves high QoS with very few reconfigurations.
%
%\enlargethispage{\baselineskip}
%
Under common pricing models~\cite{chun2019cloud}, the learning-based agents provide more favourable QoS--cost trade-offs than the reactive baselines by sustaining higher QoS with moderate capacity and fewer ineffective reconfigurations.

\begin{figure}[t]
    \centering
    \includegraphics[trim={0 0 0 10mm}, clip, width=0.95\linewidth]{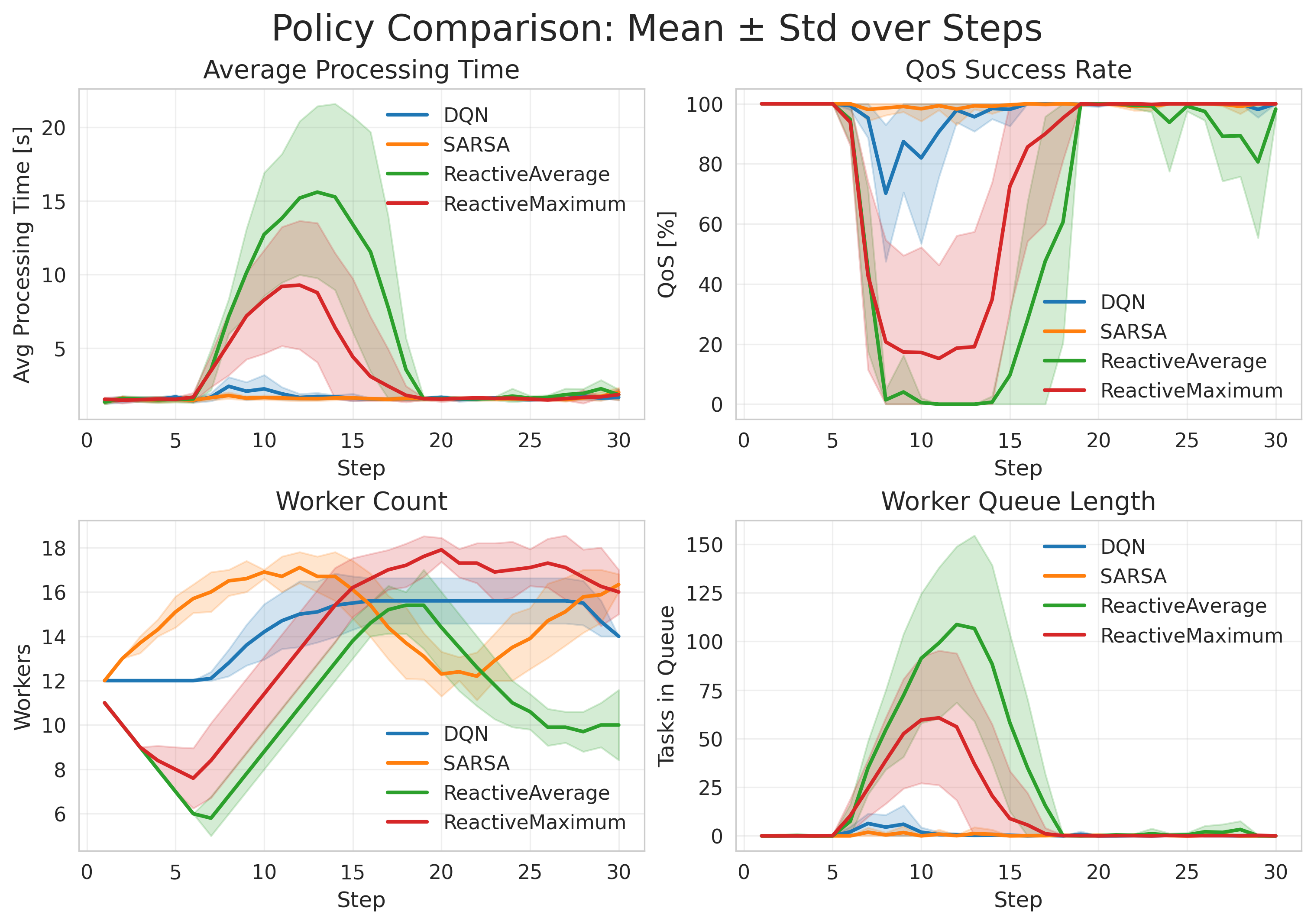}
    \caption{Policy comparison: SARSA and DQN vs. reactive baselines, showing mean$\pm\sigma$ of $\bar{T}_{proc}$, running QoS, instantaneous worker count, and worker-queue length $Q_\textrm{worker}$ over episode steps.}
    \label{fig:comparison_plots}
\end{figure}

%Viewed through the cost models of Appendix~\ref{app:cost_models} (Eqs.~\ref{eq:cost_paygo}--\ref{eq:cost_sub}), inspired by cloud pricing schemes such as those discussed by Chun et al.~\cite{chun2019cloud}, these QoS and scaling patterns place both learning-based agents in favourable regions of the QoS--cost trade-off compared to the reactive baselines; Appendix~\ref{app:cost_models} provides a more detailed comparison across pricing regimes.
%
% Moved to appendix with the formulations of cost model
\begin{comment}
Viewed through the cost models of Section~\ref{sec:formulation}, these results indicate that both learning-based agents occupy favourable regions of the QoS–cost trade-off. Under a pay-as-you-go scheme (Eq.~\ref{eq:cost_paygo}), SARSA and DQN achieve substantially higher QoS than the reactive policies at comparable or slightly higher mean worker counts, while reducing the number of ineffective scaling actions that contribute to orchestration overhead. Under a subscription-style model (Eq.~\ref{eq:cost_sub}), the similar peak worker counts across all methods imply that the reserved capacity term is comparable, but the learning-based agents use this capacity more efficiently: SARSA by keeping QoS close to 100\% with moderate scaling activity, and DQN by trading a small QoS loss for very few reconfigurations, which would translate into lower burst and scaling charges in scenarios where control-plane operations are costly.
\end{comment}
%
Finally, Fig.~\ref{fig:comparison_plots} shows the temporal evolution of the policies across the four workload phases. Both learning‑based agents maintain high QoS where the reactive baselines collapse during high‑load or rapidly varying phases despite aggressive scaling, confirming that simple threshold‑based rules are insufficient to capture the temporal structure of the workload when the action range is constrained. Overall, these results demonstrate that model‑free reinforcement learning yields autoscaling policies that dominate reactive heuristics in the QoS–cost–stability space: SARSA provides the strongest QoS guarantees, while DQN offers a nearby operating point characterised by very low scaling activity and slightly relaxed QoS targets.

%\enlargethispage{\baselineskip}

\section{Conclusions}\label{sec:conclusions}

This work presented an OpenFaaS-based implementation of a structured-parallel Farm skeleton together with an external autoscaling environment, formalised autoscaling as a Markov decision process with QoS-aware deadlines for an image-processing workload, and introduced a shaped reward that trades off service quality, backlog, worker usage, and scaling stability. We evaluated tabular SARSA($\lambda$) and Double DQN controllers against reactive baselines in a single-tenant deployment of a single Farm skeleton, using calibrated sleep-based service-time emulation.
Experiments show that learning-based policies consistently outperform reactive thresholds: SARSA attains near-perfect deadline satisfaction with moderate worker usage and fewer reconfigurations, while DQN trades a small QoS degradation for very few scaling actions and high stability. Future work will address further validation by evaluating multi-tenant settings, compositions of multiple interacting skeletons, actual workloads versus sleep-based emulation, and more challenging arrival processes including different distributions, impulse stress tests and real-world traffic traces. % We also plan to reduce the impact of OpenFaaS scale-up latency (e.g., warm pools, pre-provisioned queue-worker capacity, and predictive scaling) and to explore alternative learning paradigms and richer cost models.  \penalty 10000

\bibliographystyle{IEEEtran}
%\bibliography{temp}
\bibliography{temp}

\end{document}